\documentclass[runningheads]{llncs}
\usepackage[T1]{fontenc}
\usepackage{graphicx} 
\usepackage{hyperref}

\usepackage{booktabs}   
\usepackage{subcaption} 
\usepackage[ruled,linesnumbered,onelanguage]{algorithm2e} 

\SetCommentSty{mycommfont}
\usepackage{listings}
\usepackage{todonotes}
\usepackage{xcolor}
\usepackage{pgfplots}
\usepackage{multicol}
\usepackage{pifont}

\usepackage{enumitem}

\usepackage{balance}
\usepackage{amsmath}
\usepackage{amssymb}
\usepackage{amsfonts}

\usepackage{csquotes}

\usepackage{booktabs}   
\usepackage{subcaption} 

\newcommand{\sym}{\mathrm}
\newcommand{\matchr}{{\kern 0 em \rightarrow\,}}

\newcommand{\code}[1]{\ensuremath{\mathtt{#1}}}

\newcommand{\meth}[2][blue]{%
\if\instring{(}{#2}%
\mintinline{java}{#2}%
\else%
\texttt{\footnotesize{{\color{#1}#2}}}%
\fi
}

\newcommand\tuple[1]{\langle #1 \rangle}

\newcommand\setof[1]{\{#1\}}

\newcommand\vt{\ensuremath{\mathtt{true}}}
\renewcommand\vt{\ensuremath{\top}}

\newcommand\rw{\mathrm{rw}}

\newcommand{\tmon}[0]{\mathbf{t}\text{-}\mathbf{Mon}}

\newcommand{\mactions}{\ensuremath{\mathbb {A}}}

\newcommand{\mactionssync}{\ensuremath{\mathsf {SA}}}

\newcommand{\mevents}{\ensuremath{\mathcal {E}}}

\newcommand{\mpart}{\ensuremath{\mathrm{na}}}

\newcommand{\ordertr}{\ensuremath{\xrightarrow{\rm tr}}}

\newcommand{\orderexec}{\ensuremath{\xrightarrow{\rm e}}}

\newcommand{\mm}{\ensuremath{\mathit{e}}}

\newcommand{\rcor}{\ensuremath{\mathrm{snd}}}
\newcommand{\rfaith}{\ensuremath{\mathrm{faith}}}

\newcommand{\mtraceactions}{\ensuremath{\Sigma}}

\newcommand{\necctr}{\ensuremath{\mathrm{tno}}}

\newcommand{\mtr}{\ensuremath{\mathit{t}}}

                        \usepackage{tikz}
\usetikzlibrary{shapes,fadings,shadows,trees,matrix,intersections,patterns,decorations.markings,decorations.pathreplacing,arrows,automata,calc,decorations,fit,positioning,backgrounds}
\usetikzlibrary{shapes.geometric,decorations.pathmorphing}

\tikzstyle{marked state}=[state,draw=red!75,fill=red!20]
\tikzstyle{green state}=[state,fill=green!10]

\tikzset{
	aut/.style      = {auto,node distance=1cm,line width=2pt,>=to,thick,align=center,shorten >=1pt},
	location/.style = {state,fill=white!89!black,draw=white!50!black,minimum size=0.7cm,inner sep=0cm},
}

\tikzset{
	accept/.style  = {double},
	reject/.style  = {pattern=north east lines, pattern color=black!30},
  vtrue/.style  = {double,fill=green!60!black, text=white},
  vfalse/.style  = {fill=red!60!black, text=white, dotted, draw=black},
  vna/.style  = {fill=yellow!60!black, text=white},
  vt/.style  = {vtrue},
  vf/.style  = {vfalse},
  fit1/.style={black, rectangle, draw, solid, thin},
  smallarrow/.style={thin, -latex}
}

\tikzset{
	atomic/.style	= {minimum size=6.5cm},
	a-big/.style	= {minimum size=7.5cm},
	composite/.style= {minimum size=2.0cm, minimum height=1cm},
	explode/.style	= {node distance= 3.5cm}
}

\newcommand\bubble[3][]{%
  \node[box,#1] (#2) [#3]{};
}%
\newcommand\lbubble[4][]{%
  \node[box,#1] (#2) [#3]{#4};
}%

                       \newcommand{\secref}[1]{Sec.~\ref{#1}}
\newcommand{\defref}[1]{Def.~\ref{#1}}

\newcommand{\figref}[1]{Fig.~\ref{#1}}
\newcommand{\algref}[1]{Algorithm~\ref{#1}}

\newcommand{\cmark}{\ding{51}}%
\newcommand{\xmark}{\ding{55}}%

\newcommand{\squishlist}{
\begin{list}{-}
 { \setlength{\itemsep}{0pt}
    \setlength{\parsep}{1pt}
    \setlength{\topsep}{1pt}
    \setlength{\partopsep}{0pt}
    \setlength{\leftmargin}{0.6em}
    \setlength{\labelwidth}{1.5em}
    \setlength{\labelsep}{0.4em} } }
\newcommand{\squishend}{
 \end{list}  }

\newcommand{\cblock}[1]{}



\newcommand{\resource}{resource}

\newcommand{\threadid}{tid}

\newcommand{\id}{id}

\lstdefinestyle{JavaStyle}{
    backgroundcolor=\color{backgroundColour}, 
  language=Java,
  tabsize=2,
  breaklines=true,
  breakatwhitespace=true,
  escapechar=|*,
  numbers=left,                    
  numbersep=5pt,
  stepnumber=1,
  basicstyle=\fontsize{8.5pt}{10pt}\ttfamily,
  aboveskip=\baselineskip,
  captionpos=b,
  frame=single,
  columns=fullflexible,
  showstringspaces=false,
  extendedchars=true,
  breaklines=true,
  showtabs=false,
  showspaces=false,
  identifierstyle=\ttfamily,
  keywordstyle=\color[rgb]{0.498,0.0,0.333},
  stringstyle=\color[rgb]{0.165,0.0,0.999},
  commentstyle=\color[rgb]{0.247,0.498,0.372},
  morekeywords={String}
}

\lstdefinestyle{UseCase}{
	style=JavaStyle,
	captionpos=b,
	float=h,
	frame=none,
	backgroundcolor=\color{white},
 tabsize=100
}
\begin{document}
\title{Sound Concurrent Traces for Online Monitoring Technical Report}
%
%

\author{
  Chukri Soueidi\orcidID{0000-0002-6112-9946} \and
  Yli\`es Falcone\orcidID{0000-0002-0114-0641}}
\authorrunning{Chukri Soueidi and Yli\`es Falcone}
\institute{
\vspace*{-.3cm}
Univ. Grenoble Alpes, Inria,
CNRS, Grenoble INP, LIG,
38000 Grenoble, France\\
{\scriptsize \email{\{chukri.a.soueidi,ylies.falcone\}@inria.fr}
}
}

%
\maketitle              
\begin{abstract} 
  Monitoring concurrent programs typically rely on collecting traces to abstract program executions. However, existing approaches targeting general behavioral properties are either not tailored for online monitoring, are no longer maintained, or implement naive instrumentation that often leads to unsound verdicts. We first define the notion of when a trace is representative of a concurrent execution. We then present a non-blocking vector clock algorithm to collect sound concurrent traces on the fly reflecting the partial order between events. Moreover, concurrent events in the representative trace pose a soundness problem for monitors synthesized from total order formalisms. For this, we extract a causal dependence relation from the monitor to check if the trace has the needed orderings and define the conditions to decide at runtime when a collected trace is monitorable. We implement our contributions in a tool, FACTS, which instruments programs compiling to Java bytecode, constructs sound representative traces, and warns the monitor about non-monitorable traces. We evaluate our work and compare it with existing approaches.
\end{abstract}

  \section{Introduction}

Guaranteeing the correctness of concurrent programs often relies on dynamic verification which requires observing and abstracting the program behavior.
%
%
%
Abstraction is typically provided by \emph{traces} that contain the executed actions which can be analyzed online or post-mortem.
Traces serve as a model for property-based \emph{detection} and \emph{prediction} techniques which choose their trace collection approaches differently based on the class of targeted properties.
%
Some approaches target generic classic concurrency errors such as data-races~\cite{RVPredict,FastTrack}, deadlocks~\cite{10.1007/11678779_15}, and atomicity violations~\cite{Atomizer,WangStoller,10.1145/3373376.3378475}.
Other techniques target general behavioral properties; those are typically order violations such as null-pointer dereferences~\cite{NPE}, and typestate violations~\cite{SenPredictive,GPREDICT,SerbanutaCR12} and more generally runtime verification~\cite{leucker_brief_2009,falcone_tutorial_2013}.
%
%

When properties require reasoning about concurrency in the program, causality between events must be established during trace collection.
Data race detection techniques for instance require it to check for concurrent accesses to shared variables; as well as predictive approaches targeting behavioral properties such as~\cite{JPredictor,SenPredictive,SerbanutaCR12,RVPredict} in order to explore other feasible executions.
Causality is best expressed as a partial order over events, which are compatible with various formalisms for the behavior of concurrent programs such as weak memory consistency models~\cite{adve_shared_1996,ahamad_causal_1995,manson_java_2005}, Mazurkiewicz traces~\cite{Mazurkiewicz86,GastinK10}, parallel series~\cite{LodayaW01}, Message Sequence Charts graphs~\cite{MeenakshiR04}, and Petri Nets~\cite{NielsenPW81}.
However, while a concurrent program behaves non-sequentially, trace collection is sequential and the causality between events must be established after observation.
For many monitoring approaches, the program is a black box and the trace is the sole system information provided.
Thus, providing traces with correct and sufficient ordering information is necessary for sound and expressive monitoring.
Any \emph{representative} trace for a concurrent execution must only include correct information and must reflect all the ordering information needed to verify a behavioral property. Thus, it is best modeled as a partial order over program actions.
This paper is concerned with providing representative traces for existing verification techniques to soundly monitor concurrent programs \emph{online}.
We are interested in general \emph{behavioral} properties targeting violations that cannot be traced back to classical concurrency errors.
%
These properties are usually expressed using total-order formalisms such as LTL and finite-state machines~\cite{10.1145/93385.93442,Patterns}.
One example is a precedence property, which specifies that a resource can only be granted (event $g$) in response to a request (event $r$), and can be expressed in LTL as $\neg g \ \boldsymbol{\mathrm{W}} \ r$.
Unfortunately, these properties have received less attention in the literature in the context of concurrent programs.
Existing approaches that can monitor such properties are either not tailored for online monitoring, no longer maintained, or implement naive instrumentation.
Let us first consider dynamic \emph{predictive} techniques that target behavioral properties, such as~\cite{JPredictor,SenPredictive,SerbanutaCR12,RVPredict}.
These techniques often rely on vector clock algorithms to timestamp events, construct partial orders and build causal models to explore feasible executions.
%
%
%
%
Although they capture representative traces, they are expensive to use online and often intended for the design phase of the program and not in production environments (see~\secref{sec:related-work}). 
Furthermore, by serializing reads and writes in the program, they assume a sequentially consistent execution model~\cite{Art}.
For simple detection, this forces a specific schedule on the program, and for prediction, it restricts the space of possible interleavings.
Such techniques are incomplete and cannot prove the absence of property violations (they never claimed it) which is a reason why online monitoring is inevitable for many systems.
We turn to classical monitoring techniques and we find that many top existing approaches (based on the competitions on RV~\cite{FalconeNRT15,RegerHF16,bartocci_first_2017}) were initially developed to monitor single-threaded programs and later adapted to multithreaded programs.
Unfortunately, as shown in~\cite{El-HokayemF18}, these approaches problematically impose a total order on events using incorrect assumptions.
They have wrongly taken for granted that the instrumentation always yields representative traces. 
When events of interest execute in concurrent regions, these tools often produce unsound and inconsistent verdicts.
These approaches can benefit from representative traces to produce sound verdicts in the presence of concurrency. 
Moreover, their expressiveness can be extended to check concurrency-related behavioral properties as well.
For example, one might be interested to verify the correctness of the scheduling algorithms of tasks with data dependencies (e.g., ~\cite{VirouleauBGR16}).
%

%
We first revisit traces and qualify two properties that determine if they are good representatives of an execution: \emph{soundness} and \emph{faithfulness} (\secref{sec:traces}). %
Soundness holds when the trace does not provide false information about the order of events.
Faithfulness holds when the trace contains all the information about the order of events from the execution. 
\textbf{RQ1:} \emph{Can we collect a representative trace of a concurrent execution on the fly with minimal interference on the program?}
Vector clock algorithms have been employed and refined for several decades now~\cite{10.1145/564870.564897,10.1145/3373376.3378475,10.1145/1073814.1073818,1303344}.
However, very few (we know of~\cite{10.1145/3385412.3385993}) are directed towards online monitoring of behavioral properties; where a final trace consists of property-related events only.
These algorithms typically require blocking the execution; by synchronizing the instrumentation, program actions, and algorithm's processing to avoid data races~\cite{10.1145/564870.564897}.
With the quadratic bound on their runtime complexity and the coarse-grained synchronization they introduce, one might want to run such an algorithm off the critical path of the program.
%
We present a vector clock algorithm (\secref{sec:soundtraces}) that does not require blocking the execution and can run either synchronously or asynchronously with the program.
%
%
Asynchronous trace collection is ideal for scenarios where the monitoring overhead cannot be afforded and a small delay in the verdict can be tolerated.
For instance, in real-time systems where the system is expected to produce a result within a defined strict deadline~\cite{259423}.
Our algorithm constructs representative concurrent traces that are sound and very often faithful\footnote{The algorithm might miss, in some marginal cases, ordering information resulting in sound but unfaithful traces (see~\secref{sec:soundtraces}).}. 
As far as we know, it is unique in the context of monitoring behavioral properties that can run off the critical path of the execution.
%
%


\textbf{RQ2:} \emph{Is the collected concurrent trace good enough to soundly monitor a property?}
%
%
Monitoring single-threaded programs depend on instrumentation, which is assumed to be correct, to provide all relevant events~\cite{BartocciFFR18}.
For concurrent programs, we notice that monitorability with classical approaches depends also on the ordering information available in the trace (resp. the execution).
%
%
Firstly, consider for instance the precedence property seen previously.
If events $r$ and $g$ happen to execute concurrently in the program, a sound trace would have them as unordered events and a sound monitor should report a violation of the property.
However, in practice, when monitoring such property with an automaton, the partial order must be linearized before passing it to the monitor.
The linearization will produce an arbitrary order between $r$ and $g$, for instance, $r$.$g$ which would make the oblivious monitor miss the violation.
In such a case, the trace is not fit for monitoring the property and the monitor should be warned.
%
%
Secondly, in certain scenarios due to some partial instrumentation or logging failure, some synchronization actions might be missing from the trace, resulting in unfaithful traces where some orderings cannot be established.
This also poses soundness problems to the monitor similar to the problems with lossy traces~\cite{Joshi2017RuntimeVO}.  
To handle the mentioned problems, we extract a \emph{causal dependence relation} from a given property to know which events cannot permute in a trace and check whether a trace contains enough order information (\secref{sec:monitoringrequirements}).
We then redefine trace \emph{monitorability} for concurrent executions with a necessary condition on the trace to guarantee a sound verdict when monitoring. 
If the condition is not met, we produce warnings for the monitor. 


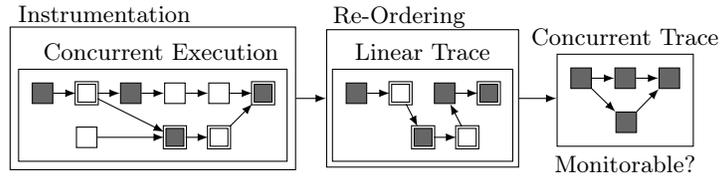
\begin{figure}[t]
	\centering
	\scalebox{1}{\begin{tikzpicture} [
  node distance=.3cm and .3cm,
  smallarrow/.style={thin, -latex},
  box/.style={rectangle,draw,minimum width=+2ex,minimum height=+2ex,inner sep=+0pt},
  RA/.style={box, fill=gray!80!black},
  SA/.style={box, double},
  hidden/.style={ draw=none},
  fit1/.style={dotted, black, rectangle, draw},
  fit2/.style={fit1, solid},
  label/.style={draw = white}]

\bubble[RA]{1t0a0}{}
\bubble[hidden]{1t1}{below=of 1t0a0}
\bubble{1t1a0}{right=of 1t1}

\bubble[SA]{1t0a1}{right=of 1t0a0}
\bubble[RA]{1t0a2}{right=of 1t0a1}
\bubble{1t0a3}{right=of 1t0a2}
\bubble{1t0a4}{right=of 1t0a3}
\bubble[SA,RA]{1t0a5}{right=of 1t0a4}

\draw[smallarrow] (1t0a0) -- (1t0a1);
\draw[smallarrow] (1t0a1) -- (1t0a2);
\draw[smallarrow] (1t0a2) -- (1t0a3);
\draw[smallarrow] (1t0a3) -- (1t0a4);
\draw[smallarrow] (1t0a4) -- (1t0a5);

\bubble[hidden]{1t1a1}{right=of 1t1a0}
\bubble[SA,RA]{1t1a2}{right=of 1t1a1}
\bubble[SA]{1t1a3}{right=of 1t1a2}

\draw[smallarrow] (1t1a0) -- (1t1a2);
\draw[smallarrow] (1t1a2) -- (1t1a3);
\draw[smallarrow] (1t0a1) -- (1t1a2);
\draw[smallarrow] (1t1a3) -- (1t0a5);

\node (exec2) [fit2, inner sep=5pt,
                          fit = (1t0a0) (1t0a5) (1t1a3)] {};
\node (labelexec2)[above left] at (exec2.north east) {{\footnotesize Concurrent Execution}};

\node (inst2) [fit2, inner sep=3pt, fit = (exec2) (labelexec2)] {};
\node (labelinst2)[above right] at ($(inst2.north west)+(0,-0.05)$) {{\footnotesize Instrumentation}};

\bubble[RA]{total1-a0}{right=3.3 of 1t0a1}
\bubble[SA]{total1-a01}{right=of total1-a0}
\bubble[RA,SA]{total1-a1}{below=of total1-a01, xshift=0.3cm}
\bubble[SA]{total1-a11}{right=of total1-a1}
\bubble[RA]{total1-a2}{right=of total1-a01}
\bubble[RA,SA]{total1-a3}{right=of total1-a2}
\bubble[hidden]{total-fit-fix}{above=0.2 of total1-a0}

\draw[smallarrow] (total1-a0) -- (total1-a01);
\draw[smallarrow] (total1-a01) -- (total1-a1);
\draw[smallarrow] (total1-a1) -- (total1-a11);
\draw[smallarrow] (total1-a11) -- (total1-a2);
\draw[smallarrow] (total1-a2) -- (total1-a3);

\node (t-total1) [fit2, inner sep=5pt, fit = (total1-a0) (total1-a1) (total1-a3) ]{};
\node (label-t-total)[above] at (t-total1.north) {\footnotesize Linear Trace};


\node (inst11) [fit2, inner sep=2pt, fit = (t-total1) (label-t-total)] {};

\node (labelinst1)[above left] at  ($(inst11.north east)+(-0.6,-0.1)$) {{\footnotesize Re-Ordering}};

\draw[smallarrow] let \p1 = (inst11.west), \p2 = (inst2.east)
    in (\p2) -- (\x1,\y2);

\bubble[RA]{partial-a0}{right=2.7 of total1-a0, yshift=0.2cm}
\bubble[RA]{partial-a1}{below=of partial-a0, xshift=0.6cm}
\bubble[RA]{partial-a2}{right=of partial-a0}
\bubble[RA]{partial-a3}{right=of partial-a2}

\draw[smallarrow] (partial-a0) -- (partial-a1);
\draw[smallarrow] (partial-a0) -- (partial-a2);
\draw[smallarrow] (partial-a1) -- (partial-a3);
\draw[smallarrow] (partial-a2) -- (partial-a3);

\node (t-partial) [fit2, inner sep=5pt, fit = (partial-a0) (partial-a1) (partial-a3)]{};
\node (label-t-partial)[above] at (t-partial.north) {\footnotesize Concurrent Trace};
\node (label-t-partial)[below] at (t-partial.south) {\footnotesize Monitorable?};

\draw[smallarrow] let \p1 = (t-partial.west), \p2 = (inst11.east)
    in (\p2) -- (\x1,\y2);








\end{tikzpicture}}
	\caption{Reconstructing concurrent traces.}
	\label{fig:approach}
	\vspace{-1em}
\end{figure}

In this paper we target shared memory concurrency, however, our work can be adapted to message passing.
We implement our contributions in a tool, FACTS, which attaches to programs running on the JVM, instruments and collects sound and faithful traces, and checks the monitorability criteria.
%
%
%
We evaluate our approach and demonstrate its effectiveness in capturing concurrent traces and provide a cost estimation on the execution time overheads (\secref{sec:evaluation}).
We also implement the algorithm from~\cite{1303344}, used in~\cite{JMPaX,SenPredictive,JPredictor}, and show how our algorithm has better advantages.
%
%
 
   \section{Issues with Linear Traces}
\label{sec:problems}
%
%
\begin{figure}[t]
    \centering
    \scalebox{0.75}{\begin{tikzpicture} [
    node distance=.2cm and .5cm,
    smallarrow/.style={thin, -latex},
    box/.style={rectangle,draw,minimum width=+9ex,minimum height=+3ex,inner sep=+0pt},
    RA/.style={box, fill=gray!80!black, text=white},
    label/.style={draw = white}]
    \node[label] (t0) {Thread 0};
    \node[label] (t1) [below=of t0]{Thread 1};
    \node[box] (t0q0) [right=of t0]{$\code{r(x)}$};
    \node[box,RA] (t0q1) [right=3 of t0q0]{$\code{notify(r,x)}$};
    \node[box] (t1q0) [right=of t1, shift={(1,0)}]{$\code{g(x)}$};
    \node[box,RA] (t1q1) [right=of t1q0]{$\code{notify(g,x)}$};
    \path[->,smallarrow]
    (t0q0) edge[] node [above]{} (t1q0)
    (t1q0) edge[] node {} (t1q1)
    (t1q1) edge[] node {} (t0q1);
    \end{tikzpicture}}
    \caption{Unsound instrumentation.}
    \label{fig:tool:strict}
    \vspace*{-1em}
    \end{figure}
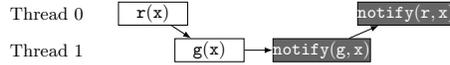

Monitoring approaches relying on total-order formalisms such as LTL and finite state machines require linear traces to be fed to the monitors as their input consists of words.
One might be tempted to immediately capture linear traces from a concurrent execution without reestablishing the causal order.
In this section, we discuss issues of collecting linear traces to motivate concurrent traces.
%
\paragraph{Advice Atomicity Assumption}
\label{sec:atomicityAssumption}

General purpose runtime verification frameworks such as Java-MOP~\cite{chen_java-mop:_2005}, Tracematches~\cite{allan_adding_2005}, and others~\cite{bartocci_first_2017} rely on instrumentation for extracting events for traces.
%
To handle concurrent programs, RV tools provide a feature to \emph{synchronize} the monitor to protect it from concurrent access and data races.
As such, threads acquire a lock before notifying the monitor and release it afterward.
However, when a program action is executed, its advice will not always execute atomically with it unless both are wrapped in a synchronization block.
%
%
 %
%
%
Consider \figref{fig:tool:strict} and the property \code{r} \emph{precedes} \code{g}. 
A context switch happens at $t1$ after the $\code{r(x)}$ is executed (but before notifying the monitor), allowing the execution of $\code{g(x)}$ and its advice before $\code{notify(r,x)}$.
It is clear how this would result in unsound monitoring.
This observation problem cannot be solved by simply checking for the absence of data races.
One needs to guarantee atomicity between all executing program actions and their advice.
However, the problem will not appear if \code{r} and \code{g} are synchronized and executed with their advice within mutually exclusive regions.
Nevertheless, we might sometimes want to capture events from concurrent regions.

 \paragraph{Forced Atomicity}
\label{sec:forcingLinearization}
One way to solve the lack of advice atomicity is to force it.
Forced atomicity can be achieved by instrumenting synchronization blocks\footnote{Such as \code{synchronized} in Java.} that wrap the program actions with their advice in mutually exclusive regions.
However, forcing atomicity introduces two new problems, one at the program level and the other at the monitor level.
First, it forces a total order between concurrent program actions; interfering with the parallelism of the program and changing its behavior. 
One needs also to minimize the area for which the lock is applied and avoid coarse-grained synchronization.
%
%
%
%
From the monitor side, the verdict will be dependent on the specific scheduling of the execution.
Take \code{r} \emph{precedes} \code{g} for instance, if these events are concurrent in the program, monitoring will produce a different verdict at each run.
Although the verdict is correct, we would like to warn the user.
Second, any information about parallel actions in the program is lost and one can no longer determine whether two actions execute concurrently initially in the non-instrumented program.
In that case, it becomes impossible to express properties on the concurrent parts of the execution.
Furthermore, RV tools mentioned above rely on AspectJ~\cite{AJ} for a high-level specification of instrumentation which is rather unfitting to instrument synchronization blocks as it cannot instrument at the bytecode level.
  \section{Characterizing a Concurrent Execution}
\label{sec:execution}

%
We first choose the smallest observable execution step done by a program, such as a method call or write operation,  and refer to this step as an \emph{action}.
%
%
%
An \emph{action} is a tuple of attribute-value pairs.
Common attributes are \emph{\id}, associated \emph{\resource} i.e. shared variable, integer \emph{\id} to distinguish different actions corresponding to the same instruction\footnote{We depict actions using the notation $\code{id.label^{\threadid}_{\resource}}$ in diagrams.}.
Actions executing in the same thread follow a \emph{thread order}.
The set of all program actions is denoted by {\mactions}.
We distinguish between two types of actions: \emph{regular actions} and \emph{synchronization actions}.
Regular actions are relevant to the property we want to monitor.
Synchronization actions, denoted by the set {\mactionssync},  are used for synchronization between different threads.
They provide \textit{release-acquire} ordering semantics~\cite{JSR133} and establish a \emph{synchronization order} on the execution.
Essentially, an acquire action $a'$ on a resource by some thread synchronizes with the latest release action $a$, if it exists, on that same resource by some other thread.
Given two threads $t$ and $u$, we highlight the following synchronization actions that can establish release-acquire:

\begin{itemize}[leftmargin=0.5cm]
    \item  $\mathit{unlock}(t,l)$/$\mathit{lock}(t,l)$: release/acquire of lock $l$ by $t$;
    \item   $\mathit{fork}(t,u)$/$\mathit{begin}(u)$: fork of $u$ by $t$/first action by $u$;
    \item  $\mathit{end}(t)$/$\mathit{join}(u,t)$: last action $t$/ $u$ blocking until $t$ ends;
    \item $\mathit{write}(t,x,v)$/$\mathit{read}(t,x,v)$: value $v$ on shared variable $x$.
    \item $\mathit{notify}(t,s)$/$\mathit{wait}(t,s)$: notify/wait a signal $s$.
\end{itemize}
The \emph{execution order}  {$\orderexec  \ \subseteq (\mactions  \times   \mactions )$} is the transitive closure of both \emph{thread} and \emph{synchronization} orders.
This order between actions is a partial order\footnote{We assume familiarity with partial orders and partially ordered sets.}, and is general enough to represent various formalisms and models of concurrent systems.

\begin{definition}[Concurrent Execution] \label{def:exec} 
    A concurrent execution is a partially-ordered set of actions $(\mactions, \orderexec)$.
\end{definition}

  \section{Sound and Faithful Concurrent Traces}
\label{sec:traces}

Ass our goal is to capture a representative trace from a concurrent execution.
In this section, we show how a representative trace preserves the properties of an execution.
We note that the trace actions $\mevents$ are often called \emph{event} in monitoring and verification approaches.

\begin{definition}[Concurrent Trace] \label{def:obexec} 
    A concurrent trace is a partially ordered set   $\mtr = (\mevents, \ordertr)$ of events such that $\mevents \subseteq  \mactions$.
\end{definition} 
We first define the notion of \emph{trace soundness}.
Informally, a trace is a sound trace if it does not provide false information about the execution.
\begin{definition}[Trace Soundness] \label{def:trace:correct}
A concurrent trace $\mtr$ is said to be a sound trace of a concurrent execution $\mm$ (noted $\rcor(\mm, \mtr)$) iff (i) $\mevents \subseteq \mactions$ and (ii) $\ordertr \subseteq \orderexec$.
\end{definition}
\noindent
To be sound, a trace (i) should not capture an action not found in the execution, and (ii) should not order unrelated actions.

While a trace that does not provide incorrect information about the execution model leads to sound monitoring, a trace can still not provide enough information about the order (for a monitor to determine a verdict).
Faithfulness is similar to \emph{completness}; it is expensive as it requires capturing all relevant causality in the program.
Informally, a \emph{faithful} trace contains all information on the order of events that occurred in the execution model.
\begin{definition}[Trace Faithfulness] \label{def:trace:faith}
A concurrent trace $\mtr$ is said to be faithful to an execution $\mm$ (noted $\rfaith(\mm, \mtr)$) iff  $\ordertr \supseteq (\orderexec \cap \, \mevents \times \mevents)$.
\end{definition}

\vspace*{-1em}
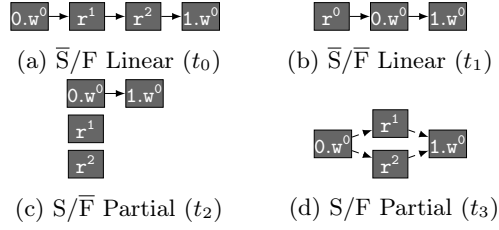
\begin{figure}[t]  
    \centering
    \subfloat[][$\mathrm{\overline{S}/{F}}$  Linear ($t_0$)\label{fig:prop1:abc}]{\makebox[0.25\textwidth]{\scalebox{0.9}{\begin{tikzpicture} [
  node distance=.3cm and .3cm,
  smallarrow/.style={thin, -latex},
  box/.style={rectangle,draw,minimum width=+3.8ex,minimum height=+3ex,inner sep=+0pt},
  RA/.style={box, fill=gray!80!black, text=white},
  SA/.style={box, double},
  hidden/.style={ draw=none},
  so/.style={dash dot},
  fit1/.style={dotted, black, rectangle, draw},
  fit2/.style={fit1, solid},
  redundant/.style={opacity=0.4},
  label/.style={draw = white}]

\lbubble[RA]{0a0}{}{$\code{0.w^0}$}
\lbubble[RA]{0a1}{right=of 0a0}{$\code{r^1}$}
\lbubble[RA]{0a2}{right=of 0a1}{$\code{r^2}$}
\lbubble[RA]{0a3}{right=of 0a2}{$\code{1.w^0}$}

\draw[smallarrow] (0a0) -- (0a1);
\draw[smallarrow] (0a1) -- (0a2);
\draw[smallarrow] (0a2) -- (0a3);

\end{tikzpicture}}}}\hspace{0.5cm}
     \subfloat[][$\mathrm{\overline{S}/\overline{F}}$ Linear ($t_1$) ]{\makebox[0.25\textwidth]{\scalebox{0.9}{\begin{tikzpicture} [
  node distance=.3cm and .3cm,
  smallarrow/.style={thin, -latex},
  box/.style={rectangle,draw,minimum width=+3.8ex,minimum height=+3ex,inner sep=+0pt},
  RA/.style={box, fill=gray!80!black, text=white},
  SA/.style={box, double},
  hidden/.style={ draw=none},
  so/.style={dash dot},
  fit1/.style={dotted, black, rectangle, draw},
  fit2/.style={fit1, solid},
  redundant/.style={opacity=0.4},
  label/.style={draw = white}]

\lbubble[RA]{0a0}{}{$\code{r^0}$}
\lbubble[RA]{0a1}{right=of 0a0}{$\code{0.w^0}$}
\lbubble[RA]{0a2}{right=of 0a1}{$\code{1.w^0}$}

\draw[smallarrow] (0a0) -- (0a1);
\draw[smallarrow] (0a1) -- (0a2);

\end{tikzpicture}}}\label{fig:prop1:linearsound}}\\
      \subfloat[][$\mathrm{{S}/\overline{F}}$  Partial ($t_2$)\label{fig:prop1:partialunfaith}]{\makebox[0.25\textwidth]{\scalebox{0.9}{\begin{tikzpicture} [
  node distance=.3cm and .3cm,
  smallarrow/.style={thin, -latex},
  box/.style={rectangle,draw,minimum width=+3.8ex,minimum height=+3ex,inner sep=+0pt},
  RA/.style={box, fill=gray!80!black, text=white},
  SA/.style={box, double},
  hidden/.style={ draw=none},
  so/.style={dash dot},
  fit1/.style={dotted, black, rectangle, draw},
  fit2/.style={fit1, solid},
  redundant/.style={opacity=0.4},
  label/.style={draw = white}]

\lbubble[RA]{0a0}{}{$\code{0.w^0}$}
\lbubble[RA]{0a1}{right=of 0a0}{$\code{1.w^0}$}
\lbubble[RA]{0a2}{below=0.1 of 0a0}{$\code{r^1}$}
\lbubble[RA]{0a3}{below=0.1 of 0a2}{$\code{r^2}$}

\draw[smallarrow] (0a0) -- (0a1);

\end{tikzpicture}}}}\hspace{0.5cm}
     \subfloat[0][$\mathrm{S/F}$ Partial ($t_3$)\label{fig:prop1:partialfaith}]{\makebox[0.25\textwidth]{\scalebox{0.9}{\begin{tikzpicture} [
  node distance=.3cm and .3cm,
  smallarrow/.style={thin, -latex},
  box/.style={rectangle,draw,minimum width=+3.8ex,minimum height=+3ex,inner sep=+0pt},
  RA/.style={box, fill=gray!80!black, text=white},
  SA/.style={box, double},
  hidden/.style={ draw=none},
  so/.style={dash dot},
  fit1/.style={dotted, black, rectangle, draw},
  fit2/.style={fit1, solid},
  redundant/.style={opacity=0.4},
  label/.style={draw = white}]

\lbubble[RA]{0a0}{}{$\code{0.w^0}$}
\lbubble[RA]{0a1}{right=of 0a0, yshift=+0.3cm}{$\code{r^1}$}
\lbubble[RA]{0a2}{right=of 0a0, yshift=-0.3cm}{$\code{r^2}$}
\lbubble[RA]{0a3}{right=of 0a1, yshift=-0.3cm}{$\code{1.w^0}$}

\draw[smallarrow,so] (0a0) -- (0a1);
\draw[smallarrow,so] (0a0) -- (0a2);
\draw[smallarrow,so] (0a1) -- (0a3);
\draw[smallarrow,so] (0a2) -- (0a3);

\end{tikzpicture}}}}
    \caption{ Four different collected traces from the execution of 1-Writer 2-Readers. }
    \label{fig:trace1:props}
    	 \vspace*{-1em}
\end{figure}

\begin{example}[Soundness and Faithfulness] \label{ex:props1}
  \figref{fig:prop1:exec} shows an execution of a Java program with three threads, a writer, and two readers\footnote{We assume that the execution follows the Java memory consistency model with the standard lock semantics}.
We choose reads and writes for simplicity, however, these can be other actions such as method calls.
An arrow connects actions ordered by the thread order and the dashed are by the synchronization order.
We omit the description of the actions labeled $\code{i}$ and $\code{d}$; these are responsible for allowing multiple readers to read concurrently.
The action labels $\code{l}$, $\code{u}$, $\code{w}$, $\code{r}$ indicate:
lock, unlock, write, read.
The lock id $\code{s}$ is the main service lock.
%
For a behavioral property such as ``no read or write happen concurrently with another read or write'', we are interested in the actions $\code{r}$ and $\code{w}$.
The order relative to these events in the execution is $\rightarrow_{e_0} = \setof{
\tuple{\code{0.w^0}, \code{r^1}},
\tuple{\code{0.w^0}, \code{r^2}},
\tuple{\code{r^1},\code{1.w^0}},
\tuple{\code{r^2},\code{1.w^0}},
\tuple{\code{0.w^0},\code{1.w^0}} 
}$. 
\figref{fig:prop1:abc} presents a linear trace of the execution  $t_0$ as captured by Java-MOP using advice atomicity assumption, see \secref{sec:problems}.
One can see that the order is $\rightarrow_{t_0} = \rightarrow_{e_0} \cup \setof{
\tuple{\code{r^1},\code{r^2}}
}$. We notice that we have faithfulness ($\rfaith(e_0, t_0)$) as $ \rightarrow_{e_0} \subset \rightarrow_{t_0}$.
However we do not have soundness ($\neg\rcor(e_0, t_0)$) as the pair $\tuple{\code{r^1}, \code{r^2}} \not\in \rightarrow_{e_0}$.
Indeed, the reads happened concurrently.
\figref{fig:prop1:linearsound} shows a trace that is neither sound nor faithful.
\figref{fig:prop1:partialunfaith} the trace captures thread order only. It is a sound trace, as it only contains $\tuple{\code{0.w^0}, \code{1.w^0}}$, and therefore no wrong information.
However, it is not faithful, as it is missing order information.
\figref{fig:prop1:partialfaith} presents a partial trace of the execution $t_3$ that is both sound and faithful.
Ideally, $t_3$ is the smallest concurrent trace collected to verify behavioral properties on reads and writes.
\end{example}
A property can be used to define the set of correct concurrent executions.
For the verification of the temporal behavior of programs, the semantics of matching properties is applied on traces~\cite{manna_temporal_1995}.
Effectively, at the semantic level, a property partitions the set of all executions into correct and incorrect ones.
Consequently, a property $\varphi(\mtraceactions)$ defined over $\mtraceactions \subseteq \mactions$  is a set of partial orders over $\mtraceactions$.

\begin{definition}[Property Satisfaction] \label{def:prop}
A concurrent execution $(\mactions, \orderexec)$ satisfies a property $\varphi(\mtraceactions)$ (noted $ (\mactions, \orderexec) \models \varphi(\mtraceactions)$)
iff $(\orderexec \cap \mtraceactions \times  \mtraceactions) \in \varphi(\mtraceactions)$.
\end{definition}
To check that a property has been satisfied, we simply ``project'' the order $\orderexec$ on $\mtraceactions$, that is, we restrict our information about the execution to $\mtraceactions$.
We then check if the projection belongs to the set of correct concurrent executions ($\varphi(\mtraceactions)$).

While our goal is verifying properties on the full execution of a program, we generally gather a subset of it as a trace.
As such, we are interested in the fact that verifying a property on the trace holds the same as it would on the full execution.
By construction (from \defref{def:trace:correct} and \ref{def:trace:faith}), we notice that the projections over $\mtraceactions$ (\defref{def:prop}) of some execution $e$ and a trace $t$ for a property over variables $\mtraceactions \subseteq \mevents \subseteq \mactions$ where we have soundness ($\rcor(e,t)$) and faithfulness ($\rfaith(e, t)$) are the same.
We deduce the following theorem.
\begin{theorem}[Property Preservation] \label{th:prop}
Given a concurrent execution $\sym{e} = (\mactions, \xrightarrow{e})$, a trace  $\sym{t} = (\mevents, \ordertr)$, and a property $\varphi$ over $\mtraceactions \subseteq \mevents \subseteq \mactions$ we have:

$(\rcor(\sym{e}, \sym{t}) \land \rfaith(\sym{e},\sym{t})) \implies (\sym{e} \models \varphi(\mtraceactions)  \mbox{ iff }  \sym{t}  \models \varphi(\mtraceactions) )$.
\end{theorem}
We say that $t$ is an appropriate abstraction of $e$; $e$ and $t$ can be used interchangeably to verify properties over $\mtraceactions$.


 



\section{Obtaining Sound Concurrent Traces}
\label{sec:soundtraces} 

We here present a vector clock algorithm that constructs sound concurrent traces. 
We assume familiarity with vector clocks and refer to~\cite{Mattern88virtualtime,lamport_time_1978} for details.
%
The algorithm differs from standard algorithms in that it can run asynchronously, allowing scenarios where a delay can be tolerated. Most, if not all vector clock algorithms in the literature block the execution to process the algorithm at each event.
It also does not require advice atomicity. Instead, it just requires instrumenting release and acquire actions with \emph{before} (resp. \emph{after}) directives.
When an action $a$ is instrumented with \emph{before} (resp. \emph{after}) direction, the notification action, $\code{notify(a)}$, should execute before $a$ (resp. \emph{after}), and no other action from the same thread should execute in between. 

\subsection{The Reordering Algorithm}

\algref{alg:vectorClockAlgorithm} maintains the following data structures.
A map $\mathbb{L}: T \rightarrow V$ holding the last timestamp seen by each thread.
%
A map $\mathbb{R}$ holding the last release action per resource, a map $\mathbb{W}$ holding the last write of a value on a shared variable, and a collection $\mathbb{T}$ which represent the concurrent trace and will be populated with timestamped actions.
On each received notification action, the algorithm sets its timestamp to the latest timestamp seen by its thread (line 3).
Then, synchronization actions in $\mactionssync$ (lines 4-5) are send to \code{ReleaseOrAcquire}, except for \emph{read} or \emph{write}. 
Actions \emph{unlock}, \emph{fork}, \emph{end}  are represented with \emph{release}, the algorithm puts them in the map entry in $\mathbb{R}$ associated with their resource in $R$ (lines 11-12).  
Actions \emph{lock}, \emph{begin}, \emph{join} are represented with \emph{acquire}, the algorithm retrieves the last action that released its resource from $R$; if found~\footnote{We omit some null checks to simplify the presentation of the algorithm, however, assume that joining $e.\mathit{VC}$ with a null value does not affect it.}, their vector clocks are merged (lines 13-15). The \emph{join} is merged with the last action seen by the finished thread. 
Actions \emph{read} and \emph{write} are handled with \code{ReadOrWrite}.
They are handled with the map $\mathbb{W}$ which is indexed by a shared variable and a value.
A write (lines 17-23) is only pushed into $\mathbb{W}$ when it is not conflicting with the entry in $\mathbb{W}$ associated with its variable and value (discussed more in the following paragraph).
A read is merged with the latest write (lines 24-26).
After that, the timestamp is incremented (line 6), and map $\mathbb{L}$ is updated (line 7).
Events will be stored in the concurrent trace, and all synchronization actions will be discarded (lines 8-9).

\begin{algorithm} [t]
\SetAlgoNoLine
\DontPrintSemicolon
\SetKwFunction{update}{Strore}
\SetKwFunction{mergeAndUpdate}{Merge}
\SetKwFunction{release}{release}
\SetKwFunction{acquire}{ReleaseOrAcquire}
\SetKwFunction{rr}{read}
\SetKwFunction{ww}{write}
\SetKwFunction{other}{RegularEvent}
\SetKwFunction{cp}{copy}
\SetKwFunction{FMain}{ReceiveAction}
\SetKwFunction{rw}{ReadOrWrite}
\SetKwProg{Fn}{}{:}{}
\SetKwProg{myproc}{procedure}{}{}
 \myproc{\FMain{e}}{
 $t$ = $\mathit{tid}(e)$\;
 $e.\mathit{VC}$ = \cp{$\mathbb{L}(t)$} \;
 \uIf{$e$ is a synchronization action}{
   send $e$ to the appropriate procedure\;
   }  
$\mathit{inc}_t(e.\mathit{VC})$\;  
 $\mathbb{L}(t) = e.\mathit{VC}$\;  
 \uIf{$e$ is not a synchronization action}{
   $\mathbb{T} \leftarrow e$ \tcp*{Add $e$ to trace  $\mathbb{T}$}
   }  
}
\BlankLine
 \myproc{\acquire{$e$}}{
   \uIf{$e$ = $\mathit{release}(t,r)$}{
    $\mathbb{R}(r) = e$ \tcp*{Update $\mathbb{R}$}
    }
    \uElseIf{$e$ = $\mathit{acquire}(t,r)$}{
    $e' = \mathbb{R}(r)$ \;
    $e.\mathit{VC} = e.\mathit{VC} \sqcup e'.\mathit{VC} $ \tcp*{Merge vector clocks}
    }
 } 

 \BlankLine
 \myproc{\rw{$e$}}{
      \uIf{$e$ = $\mathit{write}(t,x,v)$}{
         $e' = \mathbb{W}(x,v) $\;
         \uIf{ ($e' = null$) $\vee$ ($e'.\mathit{VC} \leq e.\mathit{VC}$)}{ 
         $\mathbb{W}(x,v) = e$\;
         }
         \Else{
         $\mathbb{W}(x,v) = null$   \tcp*{Conflicting write}
         }
         }
         \uElseIf{$e$ = $\mathit{read}(t,x,v)$}{
         $e' = \mathbb{W}(x,v) $\;
            $e.\mathit{VC} = e.\mathit{VC} \sqcup e'.\mathit{VC} $\;
         }         
    }   
  \caption{Vector Clock Algorithm}
\label{alg:vectorClockAlgorithm}

\end{algorithm}

The \textit{Join}, \textit{Comparison} and \textit{Copy} operations of vector clocks require $\Theta(k)$ time, linear in the number of threads $k$. \textit{Increment} operations on vector clocks, retrieving and inserting elements to the maps require $O(1)$. The reordering algorithm then requires $O(n \times k)$ time for a trace of $n$ actions. 

\subsection{Correctness Discussion} 
We now show how the algorithm always produces sound traces.
After the algorithm ends, the collected linear trace is timestamped into a concurrent trace represented by $\mathbb{T}$ that contains ordered pairs $\tuple{a,b}$, such that $a.\mathit{VC} \leq b.\mathit{VC}$.
For handling \emph{release} and \emph{acquire} actions (except for reads and writes), these actions already execute in a total order in any execution.
The algorithm performs classical operations of timestamp merging for matching actions. 
However, the correctness of our algorithm is dependent on instrumentation.
Since we do not force atomicity between an action and its notification, we must instrument actions that perform vector clock merge operations i.e. \emph{acquire} actions with the \emph{after} directive and \emph{release} actions with the \emph{before} directive.
The intuition is that if a \emph{release} action is instrumented with the \emph{after} directive, a context switch between it and its advice can lead to having an unmerged consecutive \emph{acquire}.
With proper instrumentation, whenever we observe an \emph{acquire} action, we are sure that if a matching \emph{release} exists, then it has been already been captured and processed. 
When the program only uses \emph{lock} and \emph{fork} actions for synchronization, then our algorithm guarantees both soundness and faithfulness of the concurrent trace since it captures all orderings, provided that they are all instrumented and captured.

Handling \emph{reads} and \emph{writes} is more demanding if we want to run the algorithm asynchronously and do not want to force on them a total order in the execution. 
Reads and writes are instrumented with \emph{after} and \emph{before} directives, respectively, including volatile and atomic variables~\cite{JSR166}\footnote{Atomic operations such as \emph{compare-and-swap} are handled differently since we check for the result of the operation before emitting an action.}.
When observing a read $r$ on shared variable $x$ with a value $v$ (line 24), we know its matching write was processed. 
Nevertheless, if prior to $r$, two (or more) threads performed writes, $w$ and $w'$, on $x$ with the same value $v$, then we need to make sure that $w$ and $w'$ are ordered since $r$ could be reading the value written by any one of them; hence its clock cannot be safely merged with any of them.
We say the two writes on $x$, writing the same value $v$, are \emph{conflicting} when they are not ordered, i.e. $w.\mathit{VC} \nleq w'.\mathit{VC}$ and $w'.\mathit{VC} \nleq w.\mathit{VC}$.
On a write event, the algorithm always compares it with the previously observed write on the shared variable. 
If the previous write event had written the same value and the new and previous writes are not ordered (line 19), then we clear the map entry holding the last write (line 22) so that a future read cannot merge with it.
To guarantee the correctness of the matching, we push the last write on $x$ into $\mathbb{W}$ if there are no conflicting writes (line 20).
Hence, a \emph{read} will either match its write and merge, or not.
This might have consequences on the faithfulness of the trace and not its soundness.
Faithfulness will only be affected if (1) three or more threads are writing to the same variable the same value and in a concurrent region, and (2) the program relies on those writes to synchronize.
We reasonably believe that this is an infrequent case in concurrent programs.
However, for such programs, reads and writes can be instrumented to have advice atomicity; by instrumenting synchronization blocks that wrap them with their advice in mutually exclusive regions.
This will force a total order on reads and writes and there will be no need for the procedure \code{ReadOrWrite}.
They can be treated as \emph{acquire} and \emph{release} and handled with procedure \code{ReleaseOrAcquire}.

\section{Criteria For Monitorability}
\label{sec:monitoringrequirements}

In this section, we discuss the criteria for sound monitoring with automata and concurrent traces.
In~\secref{sec:traces}, we see that a sound and faithful trace represents a program execution and can be used interchangeably when verifying a property.
However, a concurrent trace (equivalently a concurrent execution) will contain unordered events; if they are concurrent.
%
%
Many monitoring approaches rely on finite state automata as they can be used for most of the specification pattern~\cite{10.1145/302405.302672,Patterns}.
Such monitors expect a total order of events as their input consists of words. 
Given that a concurrent trace is a partial order, it must be linearized before proceeding with monitoring.
%
%
A \textit{linearization} of a partial order will impose an arbitrary order between unordered events.
However, feeding the monitor with faulty orderings of events might lead to an incorrect verdict.
%
%
For the remainder of this section, we set $\Sigma$ as the set of events over which properties are specified.
We distinguish it from $\mevents$, which is the set of runtime events that will possibly be projected to events in $\Sigma$.



%

\subsection{Monitor Causal Dependence}
\label{sec:causaldependence}
 
When observing an automaton, we might find pairs of events whose order is irrelevant to its progress; they can permute without affecting the verdict.
%
%
    The \textit{causal dependence} relation $\mathcal{D} \subseteq \Sigma \times \Sigma$ is a binary relation that is anti-reflexive and symmetric.
    It contains all pairs of events whose correct order is necessary and their permutation would lead the automaton to a different state.
Hence, for all pairs of events that do not belong to $\mathcal{D}$, their permutation after a linearization (if they occur concurrently) can be safely tolerated.
 Moreover, $\mathcal{I}_D  = (\Sigma \times \Sigma)   \setminus \mathcal{D}$ is the independence relation induced by the dependence relation $\mathcal{D}$.
 For a deterministic automaton, $\mathcal{I}_D$ is a reflexive and symmetric relation.
%
%
%
We extract $\mathcal{I}_D$  by finding pairs of actions that would lead to the same state if permuted.
\begin{algorithm}[h]

  \DontPrintSemicolon
  Given DFA $\mathcal{A} = \{\Sigma, Q, \delta^*, Q_0, F  \}$\; 
  $\mathcal{I}_D : \Sigma \times \Sigma \rightarrow \{ \mathit{tt},\mathit{ff}, \mathit{?} \}$ a map, initially all set to $\mathit{?}$.\; 
  \For{$(a,b)$ $\in$ $\Sigma \times \Sigma$}{
      \lIf{$\mathcal{I}_D(a,b)$ $\neq$ $\mathit{?}$}{
          \textbf{continue} 
      }   
      \lIf{$a = b$}{
          $\mathcal{I}_D(a,a) = \mathit{tt}$;
          \textbf{continue} 
      }
      \For{$q$ in $Q$}{
          \If{ $\delta^*(q, a.b)$ $\neq$  $\delta^*(q, b.a)$}{
              $\mathcal{I}_D(a,b) = \mathcal{I}_D(b,a) = \mathit{ff} $ \;
              \textbf{continue} to main loop
          }
      }
      $\mathcal{I}_D(a,b) = \mathcal{I}_D(b,a) = \mathit{tt} $ \;
    }  
   \caption{Generate $\mathcal{I}_D$ for DFA $\mathcal{A}$ }
   \label{alg:dependencyrelation}
  \end{algorithm}

\algref{alg:dependencyrelation} extracts $\mathcal{I}_D$ from a property specified as a DFA.
The algorithm checks for all pairs $(a,b)$ in $\Sigma \times \Sigma$ if starting from any state in the automaton $a.b$ would lead to a different state from $b.a$ (lines 8-11). If that is the case, then the automaton depends on receiving both symbols in order, and the pair is added to the dependency relation. For pairs of the same symbol, it adds them into $\mathcal{I}_D$ (line 5).

Note that, the \textit{causal dependence} relation resembles trace equivalence from~\cite{TraceTheory} which constrains the allowed linearizations of a partial order.
However, here we extract the relation from the monitor itself, and it defines \emph{word} equivalence concerning an automaton, that is, the allowed permutations of letters in a word that would eventually lead to the same verdict.

\begin{example}[Monitor Causal Dependence]
	\label{ex:dependencyrelation}
  We demonstrate with two example properties.
   \textbf{P1} states that event \code{s} responds to \code{p}  between \code{q} and \code{r}. \textbf{P2} is a mutual exclusion property that states that no read or write should happen concurrently with another write.
  %
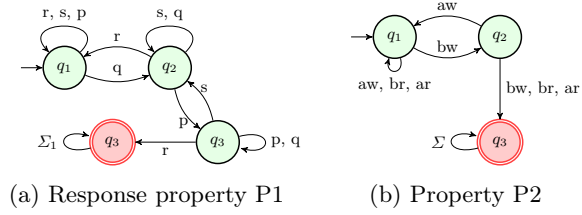
\begin{figure}[t]
    \centering
    \subfloat[][Response property P1 \label{fig:no-rw-between-w}]
    {
    	\scalebox{0.7}{\usetikzlibrary{automata, positioning, arrows}
\tikzset{
        ->,  
        >=stealth', 
        node distance=2cm, 
        every state/.style={thick }, 
        initial text=$ $, 
        }
\begin{tikzpicture}
    \node[green state] (A)                    {$q_2$};
  \node[marked state, accepting, xshift=1em]       (B) [below left of=A] {$q_3$};
  \node[green state]         (D) [right of=B] {$q_3$};
  \node[green state, initial]         (C) [left of=A] {$q_1$}; 

  \path (C) edge[loop, above, looseness=4] node{r, s, p} (C)  
        (A) edge[loop, above, looseness=4] node{s, q} (A)  
         (D) edge[loop right] node{p, q} (D)  
        (C) edge[bend right=20, above] node{q} (A)    
        (A) edge[bend right, above] node{r} (C) 
        (D) edge[below, below=0.3] node{r} (B)
        (B) edge[loop left] node{$\Sigma_1$} (B)  
        (A) edge[bend right=20, below] node{p} (D)
        (D) edge[bend right=20, above=0.3] node{s} (A);
\end{tikzpicture}


    }
    }
     \subfloat[][Property P2\label{fig:no-rw-between-rw}]
     {
     	\scalebox{0.7}{\usetikzlibrary{automata, positioning, arrows}
\tikzset{
        ->,  
        >=stealth', 
        node distance=2cm, 
        every state/.style={thick}, 
        initial text=$ $, 
        }
\begin{tikzpicture}
    \node[green state, initial] (q1) {$q_1$};
    \node[green state, right of=q1] (q2) {$q_2$};
    \node[marked state, accepting, below of=q2] (q3) {$q_3$};
    \draw   (q1) edge[loop below, looseness = 4] node{aw, br, ar} (q1)
            (q1) edge[bend right, above] node{bw} (q2)
            (q2) edge[bend right, above] node{aw} (q1)
            (q2) edge[right] node{bw, br, ar} (q3)
            (q3) edge[loop left] node{$\Sigma$} (q3);
\end{tikzpicture}}
     }
     
    \caption{Automata of Bad Prefixes.}
    \label{fig:dependency:violations}
  \vspace*{-1em}
\end{figure}
To monitor with a finite state machine monitor, the read and writes are instrumented and delimited with events \emph{bw} and \emph{aw} represent before and after a write, \emph{br} and \emph{ar} for before and after a read.
\figref{fig:dependency:violations} shows the violation automata of bad prefixes to monitor the properties.
  For each automaton, we have:

\vspace*{-0.5em}
    \begin{itemize}[leftmargin=2em]
        \item $\mathcal{I}_{D_{1}} =  \setof{\tuple{s,q}}$
         \item $\mathcal{I}_{D_{2}}  =  \setof{\tuple{br,ar}}$
    \end{itemize}
    
    From $\mathcal{I}_{D_{1}}$, we see that the monitor does not depend on the order between $s$ and $q$.
    From $\mathcal{I}_{D_{2}} $, we see that the monitor depends on the order between writes themselves, reads, and writes and reads.

\end{example}

\subsection{Trace Monitorability of Concurrent Executions}
\label{sec:tmon}
We first define the notion of necessary order for a concurrent trace, which indicates whether the trace has the needed order based on the dependence relation.

\begin{definition}[Trace Necessary Order] \label{def:tracefa} 
    We say that a concurrent trace $t = (\mevents, \ordertr)$ has the necessary orderings w.r.t. a causal dependence relation $\mathcal{D}$, noted $\mathrm{\necctr}(t,\mathcal{D})$
 
  when:
 
   $
     \ \forall \ e,e' \in \mevents :  \ \tuple{e,e'} \in \ \ordertr \ \ \vee \  \ \tuple{\mathit{label}(e),\mathit{label}(e')} \notin \mathcal{D}      
  $
  \end{definition}

\vspace*{0.5em}

  \begin{example}[Trace Necessary Order]\label{ex:traceneccessarycausality}
    Back to property \textbf{P2} from Example~\ref{ex:dependencyrelation} and the traces depicted in \figref{fig:trace1:props}.
    Trace $t_3$ is ordered enough for monitoring the property, whereas $t_2$, a trace collected considering the thread order only, does not capture order between reads and writes. As such we have:
 
 \begin{minipage}{.45\linewidth}
    \begin{itemize}
        \item[]  $\mathrm{\necctr}(t_3,\mathcal{D}_2) = \top $
    \end{itemize}
 \end{minipage}
 \begin{minipage}{.45\linewidth}
	\begin{itemize}
		\item[]  $\mathrm{\necctr}(t_2,\mathcal{D}_2) = \bot $
	\end{itemize}
\end{minipage}

     \smallskip

\end{example}

Let us recall the notion of monitorability from~\cite{Kupferman2001,BauerLTL}. 
A property $\varphi$ is \emph{monitorable}, denoted by $\mathrm{Mon}(\varphi)$, if every prefix of every trace has a finite extension that allows its monitor to reach a verdict, be it positive or negative.
Monitoring with unsound traces leads to unsound verdicts.
We redefine monitorability for concurrent programs by adding necessary conditions on the traces.

\begin{definition}[Trace Monitorability of Concurrent Executions]
  Given a property $\varphi$ with its dependency relation $\mathcal{D}$, and a trace $t$ collected from a concurrent execution $e$.
  Property $\varphi$ is monitorable with $t$, noted $\mathbf{t}\text{-}\mathbf{Mon}(\varphi)$ when
 $\mathrm{Mon}(\varphi)  \ \wedge \ \rcor(\mm, \mtr) \ \wedge \ \mathrm{\necctr}(t,\mathcal{D})$.
\end{definition}

First, the property $\varphi$ should be monitorable in the classical sense, $\mathrm{Mon}(\varphi)$, or else we will not reach a verdict.
Second, the trace should be sound, $\rcor(\mm, \mtr)$, or else we will have an unsound verdict.
Third, the trace should have all the ordering information needed by the property as per its dependency relation, $\mathrm{\necctr}(t,\mathcal{D})$, or else a linearization would produce an unsound trace.
The above indicates that a concurrent trace does not need to contain all the ordering information between events, and that the notion of faithfulness can be relaxed when monitoring.
Now, if there is missing information in a sound and faithful concurrent trace, this means that the execution itself does not contain the needed causality for monitoring the property.
As such, the user is warned about the missing order to address the problem, and a tradeoff is presented between concurrency and monitorability.
On one hand, they can synchronize the concurrent actions in the program to have them ordered, or they can force linearization of unordered actions via instrumentation as discussed in \secref{sec:forcingLinearization}.
On the other hand, they can leave the actions executing concurrently and afford inconsistent verdicts.

\subsection{Optimal Faithfulness} 
\label{sec:faithfulnessdegrees}

Since an ordering is a set of pairs, we can now define a ratio to faithfulness as the number of existing ordered pairs in a trace compared to the ordered pairs in an execution.
For a given sound trace  $\mtr = (\mevents, \ordertr)$, we define the trace faithfulness ratio as   \(\mathit{R} = |\ordertr|/|(\orderexec \cap \, \mevents \times \mevents)|\).
The faithfulness ratio cannot be greater than 1 for sound traces.
If all of the collected actions execute in synchronized regions in the program, then we get a faithful trace with $\mathit{R} =1$.

Ideally, we want to instrument programs to capture traces with the smallest optimal faithfulness ratio, denoted as $\mathrm{R}_{\varphi}$.
Such traces contain only necessary orderings for monitoring some property $\varphi$.
Moreover, $\mathrm{R}_{\varphi}$ delimits the boundary of monitorability. 
Obtaining lower faithfulness ratios leads to a non-monitorable execution.
Obtaining more than the optimal ratio means there might be a chance of optimizing instrumentation to lower the overhead on the executing program. 

\begin{property}[Degrees of Faithfulness]
  \label{prop:degrees}
  Given some faithfulness ratios $\mathrm{R}'$ and $\mathrm{R}''$ for a collected trace from an execution $e$ to monitor some property $\varphi$. 
 If $\rcor(\mm, \mtr)$ and $\mathrm{Mon}(\varphi)$ hold, we have:

    \begin{itemize}
	\item[(1)] $\mathrm{R}' < \mathrm{R}_{\varphi}   \implies  \neg \tmon(\varphi)$ 
  \item[(2)]  $\mathrm{R}'' \geq \mathrm{R}_{\varphi} \implies \tmon(\varphi) $ 
\end{itemize}
\end{property}

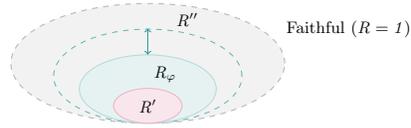
\begin{figure}[t]
  \centering
  \scalebox{0.7}{\begin{tikzpicture}
    \node[above,ellipse,minimum height=7em,minimum width=16em,draw,color=gray!60,dashed,fill=gray!10] (d) {};
    \node[above,ellipse,minimum height=5.5em,minimum width=11em,draw,color=teal!60,dashed,fill=gray!10] (c) {};
    \node[above,ellipse,minimum height=4em,minimum width=8em,draw=teal!30,fill=teal!10] (b) {};
    \node[above,ellipse,minimum height=2em,minimum width=4em,draw=purple!30,fill=purple!10] (a) {$\mathit{R}'$};   
    \path (a.north) node[above right] {$\mathit{R}_{\varphi}$}
        (b.north) node[above right=0.6cm] {$\mathit{R}''$}
        (c.north) node[right=2.3cm] { $ \ $  Faithful ($\mathit{R=1}$)};
     \draw [<->,color=teal!80] (b.north) -- (c);
\end{tikzpicture}}
  \caption{%
    Approaching optimal faithfulness.
  }
  \label{fig:dependency-onion}
\end{figure}

One can check statement (1) by simply checking if the trace satisfies the $\necctr$ condition.
As for (2), it presents us with an optimization opportunity, as we ideally want only to capture the orderings required for the monitor and not the complete orderings in the execution.
The optimization could determine the smallest set of synchronization actions (SAs) required to capture the optimal faithful ratio.
\figref{fig:dependency-onion}, shows a depiction of this optimization.
However, obtaining optimal faithfulness with instrumentation is an exciting challenge that we leave for future work.
Approaches such as sliced causality~\cite{10.5555/1770351.1770387} can be inspiring. 
Sliced causality aims to build a more relaxed causal model that allows exploring interleavings.
It does so by applying static analysis with the help of a collected trace to determine data and control dependent actions that should be kept in the causal model.
We envision an approach where multiple runs can guide instrumentation to reach $\mathrm{R}_{\varphi}$.
Nevertheless, we show an example of optimizations that can help capture fewer synchronization actions from the program.

\begin{example}[Optimizing Instrumentation]
  Let us re-examine the concurrent execution presented in \figref{fig:prop1:exec}.
  The lock associated with the counter keeping track of readers, while useful to the program, can be omitted as the additional order provided by locks and unlocks on the counter ($\code{c}$) can be obtained from the test read lock ($\code{t}$) to order reads ($\code{r}$).
  However, this would not be the case if we want to establish an order for decrements ($\code{d}$).
\end{example}

\section{Implementation}
\label{sec:implementation}

We implemented the tool FACTS (\textbf{Fa}ithful and Sound \textbf{C}oncurrent \textbf{T}race\textbf{s}) for Java programs~\footnote{The tool can be found with supplementary material.}.
The tool takes as input an instrumentation specification and a dependency relation $\mathcal{D}$ and runs attached to a running target program as a Java agent.
We discuss below the different modules.

\paragraph{Instrumentation}

Instrumentation logic is specified in separate classes written in an aspect-oriented programming style.
The user specifies the actions needed for monitoring and selects the concurrency primitives they want to collect.
FACTS extracts by default: the thread id and event name (specified by the user), the resource that is producing the event.
For monitoring actions, the user can define other static or dynamic information to extract from the program such as line number, class name, or stack values.
A \emph{scoping} feature allows specifying different filters such as package, class, field names, etc.
The tool performs instrumentation at load-time, and generated actions are passed at runtime to FACTS for processing.
%

\paragraph{Trace Reordering}
%
The vector clock algorithm processes the collected actions and pushes the timestamped monitoring actions into the concurrent trace, which is kept in memory or passed to the monitor.
Depending on the monitoring scenario, the algorithm can run on a separate thread from the executing program or in the same thread.
For representing vector clocks, we use tree clocks~\cite{treeclocks} in our implementation instead of classical vector clocks.
Tree clocks improved the overall performance of the reordering algorithm and gave us a sublinear time on join and copy operations, in contrast to classic vector clocks, which always require  $\Theta(k)$ time linear in the size of the vector.

\paragraph{Monitorability Checker}
This module checks if the concurrent trace is monitorable as discussed in \secref{sec:monitoringrequirements}. 
It runs in parallel with the reordering algorithm and is concerned only with checking the $\mathrm{\necctr}$ condition.
%
It is compatible with parametric slicing~\cite{ChenRosu}; it slices events based on the runtime information bound to them.
%
%
When receiving a processed event from the vector clock algorithm, before sending the event to the monitor, the checker checks if the event is ordered 
with the events from its slice (as per the dependency relation).
If it detects missing orderings, warnings are issued along with event names and their location in the code.

  \section{Experimentation and Evaluation}
\label{sec:evaluation}

  \subsection{Effectiveness and Cost}
We demonstrate the effectiveness of FACTS in capturing sound and faithful traces.
We measure the effectiveness and cost of obtaining sound and faithful concurrent traces.
Apart from tools dedicated to data races and atomicity violation detection, we found no available tools targeting general behavioral properties to compare with; tools that can establish causal order and instrument various custom regular actions from the program for monitoring. 
As such, we implemented the algorithm from~\cite{1303344}, used in~\cite{JMPaX,SenPredictive,JPredictor}.
The implementation uses the same vector clock data structures, tree clocks~\cite{treeclocks}, as in FACTS.

\subsubsection{Experimental Setup} 
We pick for our evaluation real-world Java applications from Renaissance~\cite{Renaissance} and DaCapo Benchmarks~\cite{blackburn_dacapo_2006}, and synthetic programs from~\cite{Art}.
We monitor properties that can be expressed with total order formalisms to demonstrate our work and show how concurrent traces can help existing monitoring approaches adapt to concurrent programs.
We compare three concurrent trace collection approaches:  with \textbf{FACTS} in both \emph{asynchronous} and \emph{synchronous} modes, and with Algorithm \textbf{A} from~\cite{1303344} which cannot be run in asynchronous mode.
We also collect linear traces as collected by Java-MOP~\cite{chen_java-mop:_2005} and show the number of ordering corrections made with concurrent traces.
We instrument the programs to collect property-related events and for concurrent traces also capture synchronization actions such as thread operations, synchronized blocks and methods, locks, reads and writes to shared variables, and spawning actors~\cite{Actors} in Akka~\cite{Akka}.

%
%
%
%


\subsubsection{Benchmarks and Specification}
The benchmarks are chosen to span different concurrency primitives such as thread operations, synchronized blocks, and methods, lock operations, reads and writes, and spawning actors~\cite{Actors} in Akka~\cite{Akka}. 

Program \textbf{akka-uct} performs load balancing of tasks using an Unbalanced Cobwebbed Tree computation with Akka~\cite{Akka}. 
Worker tasks with different priorities, \emph{urgent} and \emph{normal}, are completed using actor nodes.
We monitor a response property stating that \emph{between the submission and the execution of a task with normal priority (events $q$ and $r$ resp.) if an urgent task is submitted (event p) it should execute (event s) in between q and r}. 
The property is identical to \textbf{P1} from Ex.~\ref{ex:dependencyrelation}.
%
%
%
We have $\mathcal{I}_D = \setof{\tuple{s,q}}$ which is passed to FACTS.
We specify instrumentation to extract events from within the execution of the nodes and target node creation and send messages for synchronization actions.

Program \textbf{future-genetic} executes a genetic algorithm optimization function using Jenetics~\cite{Jenetics}. 
The program has a sequence of parallel tasks executing with contention between them.
We check \emph{whether dependent tasks execute in parallel}.
The task execution is instrumented and delimited with a \emph{before} task and \emph{after} task events ( $bt$ and $at$ respectively).
The property is a mutual exclusion property similar to \textbf{P2} from Ex.~\ref{ex:dependencyrelation} without $q_3$ and the \emph{read} events.
Here $\mathcal{I}_D$ is empty and passed to FACTS. 
We specify instrumentation to capture events that execute within the tasks and target synchronized blocks and methods and thread forks for synchronization.   

We target type-state properties~\cite{typestate} with the Dacapo benchmarks \textbf{avrora} and \textbf{fop}.
We collect traces to monitor the \text{SafeIterator} property that specifies that \emph{no thread should update a collection while another thread iterates over it}, and the \text{HasNext} property that \emph{requires calling \code{hasNext()} on an iterator before calling the \code{next()} operation}.
With FACTS, events are extracted with runtime information so that they are matched in a parametric monitoring setup~\cite{ChenRosu}.
Most of the Dacapo programs synchronize using synchronized blocks and methods, so we target those for collecting SAs.
For both properties, $\mathcal{I}_D$ is empty and all events are expected to be ordered.

We also run an implementation of the Bakery lock algorithm~\cite{10.1145/361082.361093}, \textbf{bakery (f)}.
The algorithm performs synchronization using reads and writes.
We introduce a bug such that synchronization between threads is faulty.
We check if events in the critical section are not overlapping, i.e., atomic.
We also include a classic producer-consumer program, \textbf{prods-cons (f)}, which performs synchronization with locks.
We also introduce a bug in locking consumers which would spontaneously produce events without acquiring the lock of the shared resource.
For both, we are checking a property similar to \textbf{P2} from Ex.~\ref{ex:dependencyrelation}, as such, a similar $\mathcal{I}_D$ is passed to FACTS. 
%

%
\subsubsection{Results}
\begin{figure}[t] 

  \begin{subfigure}{0.65\textwidth}
      \centering
      \includegraphics[width=1\textwidth]{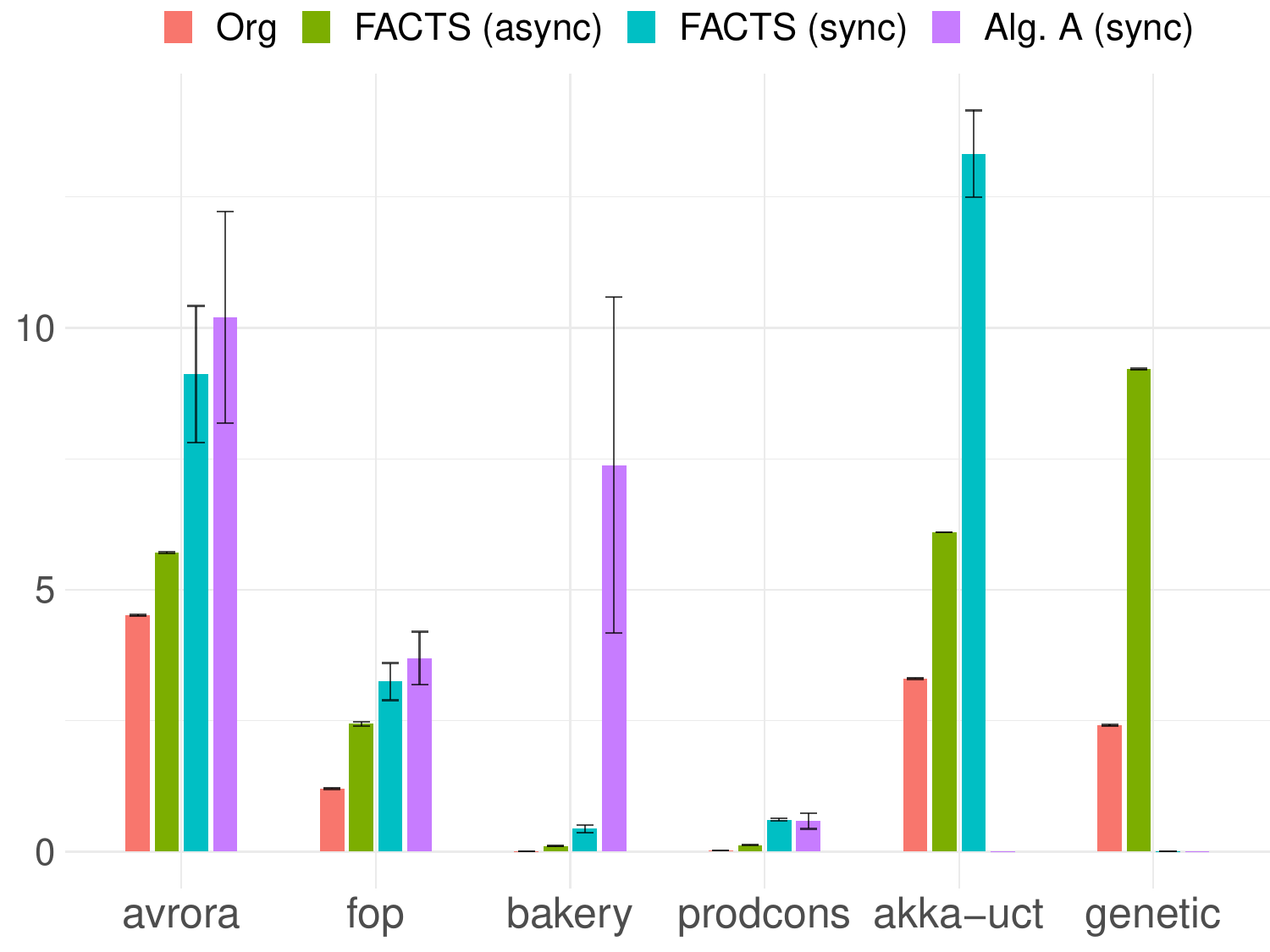}
      \caption{Execution time (s) with trace reordering.}
      \label{fig:etimes}
  \end{subfigure}
  \begin{subtable}{0.27\textwidth}
    \centering
    \begin{tabular}{rcc}
      \toprule
      &  $\tmon$   \\ \hline
     \text{akka-uct} &  \xmark    \\
     \text{future-genetic} & \cmark    \\ 
     \text{avrora}  &  \cmark   \\ 
     \text{fop} &   \cmark       \\ 
     \text{bakery}  & \cmark      \\ 
     \text{prods-cons}  & \cmark      \\ 
     \text{bakery (f)}  & \xmark      \\ 
     \text{prods-cons (f)} & \xmark      \\ \hline
     
  \end{tabular}
    \caption{Trace Monitorability.}
    \label{tab:experiment}
    \vspace*{1em}
    \centering
    \begin{tabular}{cll}
      \toprule
         \textbf{t-Mon}  &  \multicolumn{2}{c}{\textbf{Verdict}}\\
      \midrule
                 \cmark &  T: 100\% & F: 0\%\\
                  \xmark &  T: 32\% & F: 68\% \\
      \end{tabular}
      \caption{Monitoring Soundness.}
      \label{tbl:badjmop}
   
\end{subtable}%
  \caption{Experimentation Results.}
  \vspace*{-1em}
  \end{figure}

\figref{fig:etimes} report the mean execution time for 20 runs of the benchmarked programs.
We first note that FACTS was capable of producing sound and faithful traces from all benchmarks as no marginal cases (i.e. conflicting writes) were reported from our vector clock algorithm, even for the \emph{bakery} for instance which relies solely on reads and writes for synchronization.
Second, running the algorithm in asynchronous mode, FACTS (async), interferes minimally with the program as it incurs a considerably low overhead in most of the benchmarks.
Third, FACTS (sync) performs better than Algorithm A in most of the benchmarks.
We fairly believe that our algorithm interferes less with parallelism in the programs as it imposes finer-grained synchronization than Algorithm A. 
This is highlighted with the bakery algorithm which synchronizes only using shared variables. 
Algorithm \text{A} requires the update of vector clocks associated with a read or write to be atomic through synchronization (as in \secref{sec:forcingLinearization}), while our algorithm does not.
This causes the threads to spin more as more contention is added with Algorithm \text{A}.
Forth, for \textbf{future-genetic} and \textbf{akka-uct}, these programs use parallel tasks and message-passing (resp.) for managing concurrency.
We can see how capturing concurrent traces synchronously from them interferes severely with their behavior.
Algorithm \text{A} and FACTS (sync) timed out with \textbf{future-genetic}, while for \textbf{akka-uct}, Algorithm \text{A} is not intended to handle message passing.
Monitoring programs that use concurrency primitives with higher levels of abstraction need better adaption in the future; for now, we better observe and monitor them asynchronously.

Table~\ref{tab:experiment}, reports on the monitorability of the collected traces. 
The monitor is warned when $\tmon$ is false as it may produce unsound verdicts.
We introduce two buggy implementations \textbf{prods-cons (f)} and \textbf{bakery (f)}, where synchronization between threads is faulty and events 
execute without acquiring the locks, leading to missing orderings in the executions (resp. the trace).
We use both a correct implementation where $\tmon = \top$ and the faulty one where $\tmon =\bot$.
%
%
We monitor with Java-MOP~\cite{chen_java-mop:_2005} and collect the verdicts. A \emph{true} verdict (T) means the property is not violated.
We report the results of 100 executions in Table~\ref{tbl:badjmop}. 
We find that monitoring with $\tmon = \top$ yields a verdict \emph{true} (32\%) for some executions, while for others, it yields \emph{false} (68\%).
FACTS is capable of producing warnings in all executions of the faulty programs.
In the correct program, verdicts are indeed consistent because the execution of the events themselves is linearized.

  \begin{figure}[t]
    \centering
    \includegraphics[width=1\textwidth]{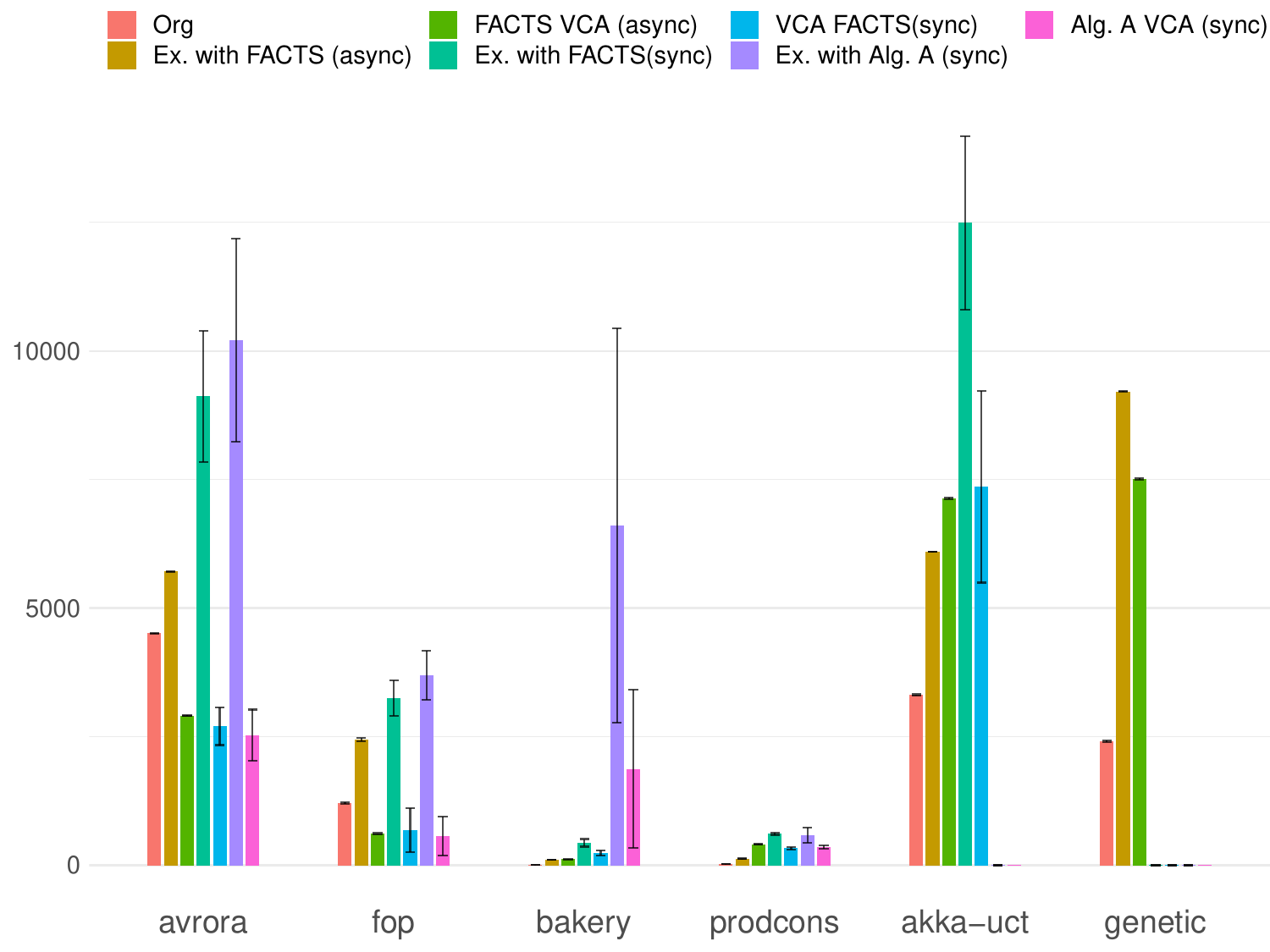}
    \caption{%
    Execution time (s) with vector clock algorithm running times.
    }
    \label{fig:vca-exec}
  \end{figure} 
  
  \figref{fig:vca-exec} reports also in addition to the execution times of the programs the execution times for the vector clock algorithms (labeled VCA). For (async) the algorithm time is excluded from the execution time.

  \bgroup
  \begin{table}[t!]
   \centering
   \caption{The table reports for each benchmark: the number of threads (Tr), execution time in seconds of the benchmark (Exec), \# of events ($\mevents$) and their type (Type: S for synchronized, P for parallel), \# of synchronization action captured ($\mactionssync$) for FACTS, vector clock algorithm time in sec (VCA),  monitorability check $\tmon$, \# of faulty pair orderings (Faulty pairs) from linear traces. K = $10^3$, M = $10^6$.}
   \renewcommand{\arraystretch}{1}
    \begin{tabular}{rrrrcrrrccr}
      \toprule
      \multicolumn{1}{c}{} & \multicolumn{4}{c}{\textbf{ORG}} & \multicolumn{4}{c}{\textbf{ FACTS}} & \multicolumn{2}{c}{\textbf{Linear Traces}} \\
      \cmidrule(lr){2-5} \cmidrule(lr){6-9} \cmidrule(lr){10-11} 
      & \multicolumn{1}{c}{Tr} & \multicolumn{1}{c}{Exec} & \multicolumn{1}{c}{|$\mevents$|} & Type & \multicolumn{1}{c}{|\mactionssync|} & \multicolumn{1}{c}{VCA} & \multicolumn{1}{c}{Exec} & $\tmon$ & \multicolumn{1}{c}{Faulty} & \multicolumn{1}{c}{Exec} \\ \hline
     \textbf{akka-uct} & 64 & 3.3 & 1.28M & P+S & 3.2M & 7.1 & 6.1 & \xmark  & 686K & 4.91 \\
     \textbf{future-genetic} & 18 & 2.4 & 17M & P+S & 1.26M & 18.1 &  9.2 & \cmark   &  - & 9.1 \\ 
     \textbf{avrora} & 8 & 4.5 & 2.5M & P+S & 3.2M & 2.9 & 5.7 &  \cmark  & - & 4.75 \\ 
     \textbf{fop} & 1 & 1.2 & 1.6M & S & 7K & 0.6 & 2.47 & \cmark   & - &  2.2  \\ 
     \textbf{bakery } & 4 & 0.1 & 400K & S & 9.3M & 7.4 & 4.5 & \cmark   & - & 0.17  \\ 
     \textbf{prods-cons } & 28 & 0.002 & 112K & S & 242K & 0.4 & 0.13 & \cmark   & - & 0.06 \\
     \textbf{bakery (f)} & 4 & 0.1 & 400K & S & 8M & 6.5 & 4.3 & \xmark   & 75K & 0.16  \\ 
     \textbf{prods-cons (f)} & 28 & 0.002 & 112K & S & 212K & 0.4 & 0.11 & \xmark   & 21K & 0.06 \\ \hline
  \end{tabular}
  \label{tab:experimentF}
  \end{table}
  \egroup

Table~\ref{tab:experimentF} reports more details about the execution.
The detected faulty pairs correspond to unordered events in the execution that are arbitrarily ordered with linear traces.
Their order differs in each execution depending on the scheduler of the execution environment.
However, they are captured correctly with concurrent traces.
The main result we have is that FACTS is capable of reporting the missing orderings and warns the monitor that the execution is not monitorable.
For \textbf{akka-uct}, urgent and normal tasks are executed concurrently. The concurrent trace is not enough and monitoring with linear traces produces unsound verdicts.
For \textbf{bakery} and \textbf{prods-cons}, the locking mechanism is faulty.
Although the events are collected from supposedly synchronized regions, one should never assume atomicity between events as one might be monitoring to check the atomicity assumption in the first place.
For \textbf{avrora}, the related events are all synchronized as they are produced by the same threads, and hence here one can use linear traces.
For \textbf{fop}, running on a single thread, linear traces are ideal and can be safely used.
For \textbf{future-genetic}, no reported faulty pairs as it seems related events are properly ordered.
The execution times with FACTS show that the cost is high when collecting representative traces.
Compared to linear traces, the overhead in the execution (Exec) is expected as FACTS is collecting extra synchronization actions.
In many cases, they are more than property events.
However, the overhead can be reduced by targeting frameworks that provide higher-level abstractions of concurrency and are built on top of low-level primitives, such as actor systems and fork-join frameworks.

  \subsection{Causal Dependence Relation in Specification Patterns}
  \begin{figure} 
      \centering 
      \includegraphics[width=.8\textwidth]{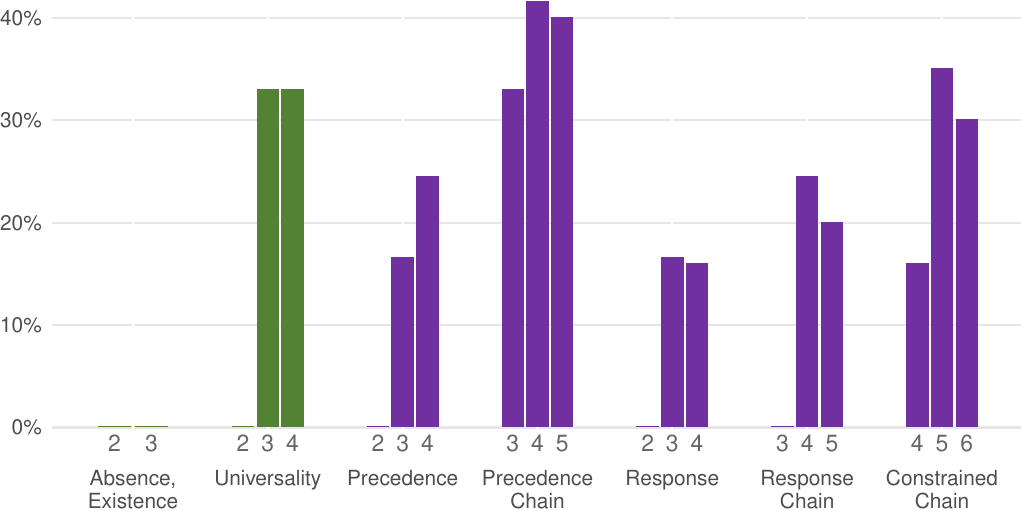}
      \caption{Percentage of pairs in $\mathcal{I}_D$ grouped by pattern and alphabet size.}
      \label{fig:IndependencePercentage} 
    \end{figure}

  We evaluate how much faithfulness can be relaxed while still being able to soundly monitor an execution.
  We extract the independence relation $\mathcal{I}_D$ of 55 event-based property specifications from~\cite{10.1145/302405.302672,Patterns} written as Quantified Regular Expressions (QREs).
  These specifications span commonly occurring patterns in the verification of programs.
  They are classified into \emph{occurrence} patterns (first two in green in \figref{fig:IndependencePercentage}) that target the occurrence of a given event, and \emph{order} patterns (in purple) that target the relative order in which multiple events occur during execution.
  An example of an \emph{order} pattern is the precedence property: some event $S$ \emph{precedes} some event $P$, which can be used to specify a requirement that a resource can only be granted in response to a request.
  
  For each specification, we generate an automaton and run \algref{alg:dependencyrelation} to find pairs in $\Sigma \times \Sigma$ with causal independence.
  We do not count pairs with the same symbol as these are in $\mathcal{I}_D$ by definition.
  For an automaton with the alphabet $\Sigma$, the total number of relevent pairs we consider is then $(|\Sigma| \times |\Sigma|) - |\Sigma|$.
  The patterns contain specifications that vary in $|\Sigma|$ ranging from 1 to 6 letters, we exclude single-letter specifications. 
      For example, for the property $s$ \emph{precedes} $p$, we find that the pair $(s,p)$ is not in $\mathcal{I}_D$. 
  However, for the property $s$ \emph{precedes} $p$ \emph{before} $r$, we find that the pair $(s,r)$ in $\mathcal{I}_D$.
  When monitoring such a property, the concurrent traces are still monitorable even when $S$ and $R$  are not ordered, hence, only requiring order between $(s,p)$ and $(p,r)$.
  \figref{fig:IndependencePercentage} reports the average percentage of pairs in $\mathcal{I}_D$ grouped by pattern and alphabet size.
  We notice higher percentages for \emph{order} patterns, which are quite common when specifying properties for concurrent programs. 
  For instance, precedence and response patterns often go together and allow users to specify properties over the order of events performed by different threads.
  The results show that most specifications contain pairs with causal independence, and that faithfulness can be often relaxed when monitoring concurrent programs.

   
\section{Related Work  } 
\label{sec:related-work}

We focus on \emph{property-based} dynamic verification techniques for concurrent programs that rely on traces.
%
More specifically, on techniques developed for monitoring behavioral properties expressed in total order formalisms and refer to~\cite{7932530} for a detailed survey.
These techniques typically analyze a trace to either \emph{detect} or \emph{predict} violations.
Detection techniques reason about single runs of a program.
We mention runtime monitoring tools, namely
Java-MOP~\cite{chen_java-mop:_2005}, Tracematches~\cite{allan_adding_2005,bodden_collaborative_2010}, MarQ~\cite{RegerCR15}, and LARVA~\cite{ColomboPS09} 
chosen from the RV competitions~\cite{FalconeNRT15,RegerHF16,bartocci_first_2017}. 
These tools allow different specification formalisms such as finite-state machines, extended regular expressions, context-free grammars, past-time linear temporal logic,
and Quantified Event Automata (QEA)~\cite{BarringerFHRR12}.
%
Detection techniques do not establish causal orderings between events and rely on trace collection approaches (discussed in \secref{sec:problems}) to order the collected events.
We have shown in this paper how this can produce unsound traces leading to unsound and inconsistent monitoring.
These tools can benefit from concurrent traces to guarantee the soundness of their verdicts.
EnforceMOP~\cite{luo-rosu-2013-issta} for instance, can be used to detect and enforce properties (deadlocks as well).
It controls the runtime scheduler and blocks threads that might cause a property violation, sometimes leading to a deadlock.
It requires forced atomicity as the scheduler needs to decide at each step if the execution on some thread continues or not.
%
%
In~\cite{DBLP:conf/ifm/CassarF16,attard_2022}, the authors present a monitoring framework for actor-based systems.
The tool detectEr monitors Erlang applications using traces collected using the native logging functionality.
Our approach targets generic concurrency primitives and can also be used with actor-based systems.
 
Predictive techniques reason about all feasible interleavings from a recorded trace of a single execution.
Their challenge is to construct sound and maximal causal models~\cite{SerbanutaCR12} that allow exploring flexibly all feasible interleavings.
%
In~\cite{1303344}, the authors present an instrumentation algorithm based on vector clocks, used in~\cite{JMPaX,SenPredictive,JPredictor} for generating the partial order from a running program.
The algorithm maintains one vector clock for each thread and two for each shared variable.
It executes synchronously with the executing program and is protected using \emph{synchronized} blocks to force an overall sequentially consistent~\cite{1675439} execution.
Vector clock algorithms typically require synchronization between the advice, program actions, and algorithm's processing to avoid data races~\cite{10.1145/564870.564897}.
Our algorithm can run synchronously or asynchronously with the program depending on the monitoring scenario.
As far as we know, it is unique in the context of online monitoring in establishing order off the critical path without the need to block the execution to process.
In~\cite{SenPredictive,2ndStrike} the work targets type-state errors.
jPredictor~\cite{JPredictor} for instance, uses sliced causality~\cite{10.5555/1770351.1770387} to prune the partial order such that only relevant synchronization actions are kept.
%
The tool is demonstrated on atomicity violations and data races; however, we are not aware of an application in the context of generic behavioral properties.
In~\cite{10.1145/2393596.2393651}, the authors present ExceptioNULL that target null-pointer exceptions.
Violations and causality are represented as constraints over actions, and the feasibility of violations is explored via an SMT constraint solver.
GPredict~\cite{GPREDICT}, for instance, targets generic concurrency properties.
It allows the user to express properties with regular expressions and provides explicit concurrency idioms such as atomic and parallel regions.
It establishes order by collecting thread-local traces and also producing constraints over actions.
In addition to being incomplete due to the possibility of not getting results from the constraint solver, the analysis from these tools might also miss some order relations between events resulting in false positives.
Of course, none of the presented predictive techniques are complete, i.e., can produce all possible feasible interleavings that lead to violations, due to the impossibility of constructing a complete causal model of the program.
Furthermore, these techniques reason on sequentially consistent execution models~\cite{Art}, restricting the space of possible interleavings of programs.
The idea is that if a property is violated in a sequential consistency, then it will surely be violated in a more relaxed execution model.
Our work focuses on providing concurrent traces for online detection techniques, and we have yet to explore their applicability in predictive contexts.
Unfortunately, many tools from the mentioned approaches~\cite{GPREDICT,JPredictor,JMPaX,SenPredictive,JPredictor} are not available.
Apart from tools dedicated to data races and atomicity violation detection, we found no available tools targeting general behavioral properties to compare with; tools that can establish causal order and instrument various custom regular actions from the program for monitoring. 
  \vspace*{-1em}
\section{Conclusion And Future Work}
\label{sec:conclusion}

We present a general approach for defining and collecting traces of concurrent programs for runtime monitoring.
We investigated the limitations of linear traces and showed how they lead to inconsistent and unsound verdicts.
We establish on the fly the causal ordering of events using a novel vector clock algorithm that does not require blocking the execution.
For monitoring frameworks relying on totally ordered traces, we redefined the monitorability of concurrent traces to avoid unsound verdicts.
We implemented FACTS to collect traces from Java programs and assessed the overhead of our approach using synthetic and real-world applications.

Collecting concurrent traces for runtime monitoring opens the path to interesting problems.
Firstly, monitoring techniques can be revisited and extended to take into account the partial order.
Tools relying on total order in traces can use concurrent traces to check if a trace has the needed causality and, if not, produce warnings.
Moreover, specifications (and formalisms) that can match traces obtained from FACTS can be elaborated to extend the expressiveness of monitoring to check concurrency-related behavioral properties.
For example, our approach is applicable for OpenMP runtimes; it can be used to verify the correctness of the scheduling algorithms of tasks with data dependencies (e.g., ~\cite{VirouleauBGR16}).
We can verify that the runtime never schedules two dependent tasks in parallel.
%
Secondly, it is now possible to define and quantify optimizations for capturing sound and faithful traces.
How to obtain optimal faithfulness with instrumentation is still an interesting challenge.
Thirdly, for scenarios where the execution does not have the causality needed between events, the user can be given facilities to preserve some safety property by enforcing the order.
Linearization of concurrent events may be achieved on the fly by JVM hot swapping.
 
Finally, we will reduce overheads induced by capturing synchronization actions.
We aim to include plugins to FACTS to instrument concurrency primitives with higher-level abstraction such as fork-join~\cite{DBLP:conf/java/Lea00} and software transactional memory~\cite{10.1145/1065944.1065952}.
Tailoring instrumentation to these frameworks reduces the number of collected synchronization actions.
It requires assumptions about the correctness of these frameworks.
These assumptions can be checked using static and dynamic analysis.
%


%
 \bibliographystyle{splncs04}
\bibliography{biblio}

\end{document}